\documentclass[
    ,final            
  ]
  {aipproc}

\layoutstyle{6x9}

\begin{document}

\title{Compact Binaries as Sources of Gravitational Radiation}


\author{M. Coleman Miller}{
  address={University of Maryland, Department of Astronomy,
College Park, MD  20742-2421, USA}
}

\begin{abstract}
With current terrestrial gravitational wave detectors working
at initial design sensitivities, and upgrades and space missions
planned, it is likely that in the next five to ten years
gravitational radiation will be detected directly from a
variety of classes of objects.  The most confidently expected
of these classes is compact binaries, involving neutron stars
or black holes.  Detection of their coalescence, or their
long-term orbits, has the potential to inform us about the
evolutionary history of compact binaries and possibly even
star formation over the past several billion years.  We
review what is currently known about compact binaries as
sources of gravitational radiation, as well as the current
uncertainties and what we expect to learn from future detections
of gravitational waves from these systems.
\end{abstract}

\maketitle


\section{Introduction}

Ground-based gravitational wave detectors are now operational and
taking data in many laboratories.  The current maximum sensitivity
is such that a double neutron star inspiral could be detected out
to $\sim 20$~Mpc, more than the distance to the Virgo cluster of
galaxies. The next generation of improvements, expected in less
than a decade, will enhance sensitivity to the point that regular
detections of many events are likely.  In addition, the planned
launch in the next decade of space-based gravitational wave
detectors such as LISA will allow a complementary view of the
lower frequency gravitational wave universe, which is inhabited by
thousands of known sources and will likely prove to be an
important and precise probe of strong gravity.

Gravitational wave sources are typically divided into four
categories: binary inspirals, continuous sources such as
spinning neutron stars, burst sources such as supernovae, and
stochastic sources such as turbulence in the early universe.
Of these categories, binaries are the best-understood astrophysically.
They have therefore received close attention, in terms of optimal
frequency ranges for detection, data analysis and waveforms,
and astrophysical rate estimates and scenarios.  

Here we discuss binary sources of gravitational waves, for both
ground-based high-frequency detectors and space-based low-frequency
detectors.  In \S~2 we discuss double neutron star binaries.  
In addition to reviewing rate estimates and
uncertainties, we discuss the information that will be obtained
about the masses and spins of neutron stars, as well as about
their evolutionary histories and their possible link to short
gamma-ray bursts.  In \S~3 we examine binaries involving a
stellar-mass black hole and either a neutron star or another
black hole.  In \S~4 we turn to more massive black holes, in
the $M>10^2\,M_\odot$ range, and their observability in binary
inspirals in either relatively local star clusters or in dark
matter halos in the crucial $z\sim 5-30$ epoch of structure
formation.  In \S~5 we survey the remarkable recent progress in
numerical relativity made by many groups, and we present our conclusions
in \S~6.

\section{NS-NS Binaries}

The discovery and observation of the double neutron star binary
PSR~1913+16 by Hulse and Taylor \cite{HT75}, plus similar
systems, demonstrates that these orbits decay at a rate that is
within 0.1\% of the prediction of general relativity. With
several such systems now known that will spiral together within
a few billion years or less, it is possible to make informed
estimates of the rate of such inspirals per Milky Way
Equivalent Galaxy, or MWEG.  Several groups have made such
calculations \cite{Phinney91,Kalogera04,dFP06}, with rates
ranging from a few to a few hundred per million years per
MWEG.  When combined with the volume expected to be probed by
second-generation ground-based detectors such as the advanced
Virgo and LIGO instruments, this suggests detection rates of
tens per year. There is still substantial uncertainty in this
number, because among other things it is necessary to guess the
number of sources below current thresholds of radio detection.

When double neutron star mergers are detected, they will yield
a number of astrophysical returns.  One obvious yield will be
a far larger number of high-precision neutron star masses than
exist currently.  Radio observations of the current handful
of binary pulsar systems suggest that the neutron stars in these
systems occupy a relatively narrow band in masses, from
$1.25\,M_\odot$ (pulsar ``B" in the PSR~J0737--3039 system;
\cite{Lyne04}) to $1.44\,M_\odot$ (the pulsar in the first
binary pulsar system PSR~1913+16; \cite{HT75}).  Other observations,
notably of the NS-WD system PSR~J0751 \cite{Nice05} and of X-ray timing
from neutron star low-mass X-ray binaries \cite{zhang98,BOM05a,
BOM05b,BOM06} suggest that some neutron stars have masses
on the order of $2\,M_\odot$, but statistical and systematic
uncertainties, respectively, prevent any definitive conclusions
at this time.  

Measurement of tens of pairs of neutron star
masses will increase the sample dramatically, although it will
still be possible that it is the evolutionary history of such
systems, rather than fundamental nuclear physics, that funnels
the masses into a narrow range.  If radius measurements are
possible along with the masses, this will place strong limits
on the properties of dense matter regardless of what the
masses are \cite{LP01}.  Neutron star angular momenta might 
also be measured via their frame-dragging effects,
but known double neutron star sources and binary
evolution theory both suggest that the dimensionless spin
$j\equiv cJ/GM^2$ will be much less than 0.1 and thus difficult
to detect.

Mergers of two neutron stars also have a prospect of being
electromagnetically bright, perhaps producing short gamma-ray
bursts (GRB; see \cite{Mes06} for a recent comprehensive
review).  If so, coincident observations of a short
gamma-ray burst and gravitational waves from the event will
establish their nature immediately.  The rate per volume is
difficult to establish due to many uncertainties, but
even a single such event would have major
implications.  It has been pointed out \cite{KP93} that if, as
expected, the gamma rays come from the orbital axis, the
strength of the observed gravitational wave signal will be
greater than average.  The same considerations apply if short
GRBs are caused instead by mergers of neutron stars with
stellar-mass black holes, which we discuss in the next
section.

\section{BH-NS and BH-BH Binaries}

No stellar-mass black holes are known in binaries with neutron
stars or other black holes.  As a result, there is a lack of
observational guidance about the expected merger rate, and indeed
the formal lower limit could be zero.  This can be turned
around to note that detections via gravitational waves will thus
open up new understanding for us about the evolutionary history of
such systems.  

Incidentally, unlike for NS-NS systems, mergers in globular
clusters might contribute significantly to the BH-NS and BH-BH
rates.  The argument against their contribution to the NS-NS rate
is straightforward (see \cite{Phinney91} for an early version). The
Galaxy has roughly 100 globular clusters.  Suppose that each
cluster has 200 neutron stars (probably an overestimate), and
that every one of them pairs with another neutron star and merges
within a Hubble time (certainly a gross overestimate).  The rate
is then $R=100\times 100/10^{10}$~years, or $R\approx
10^{-6}~{\rm yr}^{-1}$. The estimated rate in the disk per MWEG
is $\sim{\rm few}\times 10^{-5}~{\rm yr}^{-1}$ \cite{Kalogera04}, so
globulars are far down in the total rate.  Even for ellipticals,
where the specific frequency of globulars is an order of
magnitude larger than for spirals, the extreme upper limit is
still less than the rate in the main portion of the galaxy, and a
realistic value is orders of magnitude less.  In contrast, since
no BH-NS or BH-BH systems are known, it is possible that the
dominant contribution comes from the complex dynamical
interactions possible in dense stellar systems such as globulars.

Despite the uncertainty, there is optimism that the observed BH-NS
or BH-BH merger rates could be larger even than the NS-NS rates,
simply because the higher system mass allows detection out to
several times the distance of detectable NS-NS mergers, hence a 
volume of (several)$^3$ times larger. As with NS-NS mergers, if a BH-NS
merger is detected coincident with a short GRB it will settle the issue
of the central engine of these events, and will also settle the
issue of whether black holes swallow neutron stars whole or whether
they spit out enough matter to produce strong electromagnetic signals
(for analytical and numerical explorations in full general relativity, see
\cite{Miller05a,Faber06,SU06}).  Resolution of 
this issue from the theoretical side
will require a full no-approximation treatment of general relativistic
magnetohydrodynamics (MHD).  Encouragingly, independent groups have
recently written and tested such codes, so answers to these and other
strong-gravity MHD question may not be far off.  This will also help
determine whether either NS-NS or NS-BH mergers can eject 
nucleosynthetic products with unique signatures (due to the extreme
neutron richness of the star), and if so, whether those products can
enrich significantly the interstellar or even intergalactic medium
\cite{Lemoine02,PGF02,Inoue03,SM04,Rosswog05}.

\section{Black Holes Above $\sim 10^2\,M_\odot$}

Black holes above stellar mass certainly exist in the universe, but
there are many fundamental questions remaining about them.  For
example, there is a clear link between the kinematical properties
of galactic bulges and their central supermassive black hole, but
the origin of this link and its role in galaxy formation is still
being debated.  There is also strong but not yet conclusive evidence
of intermediate-mass black holes in the $\sim 10^2-10^4\,M_\odot$
range, which might be forming at the present epoch as well as at
higher redshifts.  The maximum characteristic frequency of a gravitational
wave is $\sim{\rm few}\times 10^3~{\rm Hz}(M_\odot/M)$ for the
inspiral portion, depending on the spin, and a comparable 
frequency $\sim 10^4~{\rm Hz}(M_\odot/M)$ for the ringdown.
Therefore, black holes of mass $<10^3\,M_\odot$ might be observable
with ground-based detectors (with assumed frequency range
$\sim 10-2000$~Hz), but more massive black holes can only be seen
with space-based low-frequency detectors (e.g., LISA, with a range
$\sim 10^{-5}-10^{-1}$~Hz).  We will thus separate these and discuss
supermassive black holes first, followed by intermediate-mass black
holes.

\subsection{Supermassive black holes}

For our purposes, a supermassive black hole (or SMBH) is the
central, most massive, black hole of a galaxy or dark matter halo.
The mass range is therefore large in principle, from perhaps $\sim
10^3\,M_\odot$ for halos containing the first stars to $\sim
10^{10}\,M_\odot$ in the largest galaxies today.  In the process
of hierarchical structure formation, as halos merge their central
black holes may as well, leading to LISA-detectable signals from
the $z\sim 5-30$ era of the first galaxies.  The rate and
properties of these mergers have been calculated by a large number
of researchers (e.g.,
\cite{MHN01,WL03,VHM03,ITS04,Sesana04,RW05}).  The rate estimates
span some five orders of magnitude, from a minimum of $\sim
0.1~{\rm yr}^{-1}$ to $>10^4~{\rm yr}^{-1}$, with the most recent
and comprehensive models typically in the ballpark of tens per
year.  The prime reason for such uncertainty appears to be that it
is currently unknown what halo masses can produce massive black
holes; since there are more low-mass than high-mass halos,
efficient production of black holes in the $M_{\rm
halo}<10^8\,M_\odot$ range would imply higher rates and also a
higher redshift start to such mergers.

There are additional uncertainties related to the effectiveness
of dynamical friction on halos in the early universe \cite{Taf03}, and
to the role of kicks in black hole mergers in the early
universe \cite{Mer04,BMQ04,Hai04,MQ04,YM04,Vol05,Lib05,MAS05}
and for intermediate-mass black holes
\cite{Tan00,MH02a,MH02b,MT02a,MT02b,MC04,GMH04,GMH06,OL06}.  Kicks
are an especially current topic because recent progress in numerical
relativity has yielded dramatic advances in understanding of the
strong gravity contribution to kicks (see below), which complements
more accurate post-Newtonian calculations of the kick during
the inspiral portion of coalescence.  The net result is
that LISA observations of these early-universe mergers will
encode unique information about key aspects of early structure
formation.  This is especially true because the mass range to
which LISA will be most sensitive ($\sim 10^4-10^6\,M_\odot$)
is extremely difficult to observe electromagnetically
due to the low mass and luminosity of the associated galaxies
and the small range of influence of the black holes.

\subsection{Intermediate-mass black holes}

We define intermediate-mass black holes (IMBHs) to be black holes
of mass greater than could form from a single star in the current
universe (thus $M>10^2\,M_\odot$) that are not the central black
holes of galaxies or dark matter halos (and hence probably have
masses $M<10^{4-5}\,M_\odot$; see \cite{MC04,vdm04} for
overviews).  It has been proposed that these could be the remnants
of massive Population III stars \cite{MR01}, and an already massive
black hole with initial mass $M_{\rm init}>10^2\,M_\odot$ might be
able to grow over billions of years via collisionless interactions
in dense stellar clusters 
\cite{Tan00,MH02a,MH02b,MT02a,MT02b,GMH04,GMH06,OL06}.  In my
opinion, however, the currently most promising origin for these
objects is as the result of runaway stellar collisions in young
massive stellar clusters whose relaxation time for the most
massive stars is less than their $\sim 2\times 10^6$~yr main
sequence lifetime \cite{Ebisuzaki01,PM02,PZ04,GFR04,OL06,GFR06,
FRB06,FGR06,Fregeau06}.  After becoming black holes, they may
acquire stellar companions in the cluster by exchange or tidal
interactions and the resulting accretion events could explain at
least some of the ultraluminous X-ray sources.

Mergers of stellar-mass black holes are likely to be rare enough
and distant enough that their detection rate with LISA is
discouragingly low \cite{Will04}.  However, some recent simulations
suggest that for  young dense stellar clusters with realistic
initial binary fractions $f_b>0.1$, more than one IMBH could form
in a given cluster via collisional runaways
\cite{GFR06,Fregeau06}.  If so, the IMBHs will merge with each
other within a few million years.  It has also been proposed that,
since massive young clusters are themselves clustered, a
cluster-cluster merger could lead to the mergers of their IMBHs
\cite{AF06}.  Either scenario leads to
signals potentially detectable to redshifts $z\sim 1$, which is far
enough to probe some features of active star formation that are not
easily observable in other ways \cite{Fregeau06}.

Yet another possibility is that if a cluster with an IMBH forms
close enough to the center of its host galaxy, the cluster
spirals in to the center in less than a few billion years.  The
cluster will be dissolved by the tidal forces in the galactic
nucleus, and the IMBH can then merge with the SMBH.  Such an event
would lead to an extreme mass ratio inspiral (EMRI; for example,
$\sim 10^3\,M_\odot$ with $\sim 10^6\,M_\odot$), but one with a
much stronger signal than traditional $10\,M_\odot$-$10^6\,M_\odot$
EMRIs.  This could lead to especially precise and model-independent
probes of the spacetime around rotating SMBHs.  Rate estimates are
difficult because of many uncertainties, but initial analytical
\cite{Miller05b} and numerical \cite{PZ06,Mat06} explorations
suggest that LISA rates of a few to tens per year are plausible.

\section{Progress in Numerical Relativity}

Focusing now on mergers between two black holes of comparable
mass, the coalescence process has typically been divided into
the stages of inspiral (from large separations down to the point
of dynamical instability), merger (from dynamical instability to
the overlap of horizons), and ringdown (in which the merged and
lumpy common horizon settles into a Kerr state).  The ringdown
phase has been understood for some time using perturbation
theory, and substantial analytical progress has been made on the
inspiral via post-Newtonian (e.g., \cite{Blanchet04}) and
effective one-body \cite{BD99} approaches.  The merger phase,
however, is not accessible analytically with any precision,
because it involves strong nonlinearities.  This stage can
therefore only be treated with full numerical solutions of
Einstein's equations.  This is  particularly tricky because even
though in physical terms the equations do not have a preferred
gauge, gauge choice makes a significant difference in the
stability of numerical evolutions. In addition, the presence of
a physical singularity and of the coordinate singularity at the
horizon pose significant challenges.

Prior to 2005, although formal mathematical progress had been made
in casting the equations in well-behaved ways \cite{SN95,BS99},
numerical evolutions were still tantalizingly difficult. This all
changed in the summer of 2005, with Frans Pretorius' stable
evolution of two equal-mass nonspinning black holes over two full
orbits, including the merger and ringdown
\cite{Pretorius05,Pretorius06}. Multiple other groups followed with
results in short order with their own equal-mass nonspinning
coalescences  \cite{Cam05,CLMZ05,BCCKV05,CLZ06a,BCCKV06a,VBKC06}
and initial results have now been reported for unequal-mass
nonspinning mergers \cite{HSL06,BCCKVM06b,Gonzalez06} and for mergers of
spinning equal-mass black holes \cite{CLZ06b,CLZ06c}.  It is
encouraging to note that the specifics of these numerical methods
differ significantly from each other, suggesting that many robust
paths to solution have been discovered, and that agreement between
the methods provides a strong test of the reliability of the
waveforms.  In particular, it seems that at present there is
clear understanding of the waveform of the last $\sim$orbit of the
merger of two equal-mass nonspinning black holes, based on
cross-comparison of results.   Progress continues to be rapid, and
it will be interesting to see whether numerical template banks can
be constructed that span a reasonable range of spins, orientations,
and mass ratios.

From the astrophysical standpoint, the most interesting output
from such calculations is the recoil produced by the gravitational
radiation emitted during a merger with some asymmetry (such as
unequal masses or spins).  This is particularly relevant to the
redshift $z\sim 5-30$ universe, when hierarchical merging of dark
matter halos is expected to lead to mergers of their central
massive black holes.  If the black hole mergers produce a strong
enough kick to eject the remnant from the merged halo, then the
halo will, for a time, be without a central black hole.  Recent
numerical simulations (e.g., \cite{HSL06,BCCKVM06b,Gonzalez06}) 
bolster earlier analytic
suggestions that the presence of a central black hole can
have a major influence on galactic development, hence gravitational
radiation kicks at this stage can have a key effect on galactic
evolution.

There have, therefore, been a large number of analytic calculations
of the recoil produced by radiation during the inspiral phase
\cite{Per62,Bek73,Fit83,FD84,RR89,W92}, and some recent estimates
of the contribution from the merger as well
\cite{FHH04,BQW05,DG06}.  However, since the majority of the kick
is in the strong-gravity regime, once again numerical calculations
are necessary to get values that are precise enough for
astrophysical purposes.  Progress in this realm has also been
rapid, with the first fully numerical calculations
\cite{HSL06,BCCKVM06b,Gonzalez06} suggesting that, for example, two
nonspinning black holes in a mass ratio of 1.5:1 will produce a net
kick of $\approx 90$~km~s$^{-1}$, which is consistent with analytic
calculations and is reliable to $\sim$10\%, sufficient for
astrophysical applications.  There is still a great deal to explore
about kicks; in particular, it will be necessary to map out the
dependence of kick speed on mass ratio and also on black hole spins.

\section{Conclusions}

The current rapid progress in gravitational wave instrumentation as
well as numerical relativity make it realistic to think that within
a decade detections of the gravitational radiation from compact
object inspirals will become routine, and that the results will be
interpreted with confidence in a physical and astrophysical
framework.  For some of the sources, such as double  stellar-mass
black hole mergers, there will be no electromagnetic counterpart,
and hence gravitational radiation will provide our only glimpse at
these events.  For others, such as double neutron star or NS-BH
mergers, there could well be spectacular EM counterparts such as
gamma-ray bursts, and simultaneous EM and GW detections will
provide  profound and complementary information about some of the
most luminous events in the universe.  Either way, the era of
gravitational wave detections is likely to bring unanticipated
discoveries to the realm of compact object astrophysics.

\end{document}